\documentclass[10pt, conference, compsocconf]{IEEEtran}

\ifCLASSINFOpdf
\else
\fi
\usepackage[colorinlistoftodos,prependcaption,textsize=small]{todonotes}
\usepackage{blindtext}
\presetkeys%
    {todonotes}%
    {inline,backgroundcolor=yellow}{}

\graphicspath{{.}{images/}}
\DeclareGraphicsExtensions{.pdf}

\begin{document}
%
\title{Smart Multihoming in Smart Shires:\\
Mobility and Communication Management for Smart Services in Countrysides\footnotemark}


\author{\IEEEauthorblockN{Stefano Ferretti, Gabriele D'Angelo, Vittorio Ghini}
\IEEEauthorblockA{Department of Computer Science and Engineering, University of Bologna\\
Bologna, Italy\\
\{s.ferretti, g.dangelo, vittorio.ghini\}@unibo.it}
}


%


\maketitle

\footnotetext{The publisher version of this paper is available at \url{http://dx.doi.org/10.1109/ISCC.2016.7543862}.
\textbf{{\color{red}Please cite this paper as: ``Stefano Ferretti, Gabriele D'Angelo, Vittorio Ghini. Smart Multihoming in Smart Shires:
Mobility and Communication Management for Smart Services in Countrysides. Proceedings of the IEEE Symposium on Computers and Communication (ISCC 2016), Messina (Italy), 2016.''}}}

\begin{abstract}
This paper discusses on the need to focus on effective and cheap communication solutions for the deployment of smart services in countrysides. We present the main wireless technologies, software architectures and protocols that need to be exploited, such as multihop, multipath communication and mobility support through multihoming. We present Always Best Packet Switching (ABPS), an operation mode to perform network handover in a seamless way without the need to change the current network infrastructure and configuration. This is in accordance with the need of having cheap solutions that may work in a smart shire scenario. A simulation assessment confirms the effectiveness of our approach.
\end{abstract}


%
\IEEEpeerreviewmaketitle

\section{Introduction}

Smart cities are one of the main current scientific and technological concerns in the field of ICT. There is a main interest to integrate multiple technological solutions in a way that would improve the management of cities' assets. While these innovations are of paramount importance, the side-effect of this focus on metropolitan areas is that all problems related to non-metropolitan areas, i.e.~countrysides and rural areas, are completely ignored. A consequence is that all trends show that people will move even further from countrysides toward cities \cite{perez}. This would lead to a depopulation of countrysides and to an increment of the digital divide between metropolitan and non-metropolitan areas, even in the same countries.

Non-metropolitan areas can be very different and various depending on the specific geographical location we are considering, e.g.~European countrysides are very different to those in the Americas or in Asian countries. Anyhow, while very diverse, these areas typically share a lack of innovative solutions that optimize the use of potentially available resources.  Thus, there is a need to focus on these areas, that need smart management solutions.

While the problem is wide and involves several different aspects concerned with technologies, management, logistics, in this paper we put the focus on communication technologies. The solution cannot be, for instance, adding novel wireless antennas in countrysides by itself. Neither, it is not possible implementing (costly) smart cities services to make them work in a country territory, due to the very different economic circumstances that would make these services not feasible in such contexts. There is the need for innovative, self-configuring and cheap solutions, possibly not strictly dependent to the presence of a classic networking infrastructure.

Devices should be put in the condition to optimally exploit their network interfaces and capabilities. For instance, mobile users should be able to take advantage of all possible wireless network connections, in a seamless way. This is an open issue considered also in smart cities, but in a this context, where connections might be quite intermittent, it becomes extremely important. For similar reasons, we should account for a massive use of techniques such as Mobile Ad-hoc NETworks (MANETs), Device-to-Device (D2D) communications, spontaneous, opportunistic and delay tolerant networks \cite{Bellavista:2015:MQC:2793474.2793489}. Ad hoc and multi-hop communications can be used to cover a certain area-of-interest to disseminate a content, or even to relay messages from an infrastructure-less area where an Internet connection is not available, towards an access point, so that requests and contents can be pushed (pulled) to (from) the Internet. Devices must be able to self-configure their communication strategies, using all the possible solutions offered in the area they are.

In this paper, we specifically deal with multihoming and show how effective the use of multiple networks can be in order to provide seamless communications. We developed a cross-layer software architecture, termed Always Best Packet Switching (ABPS), that uses a proxy-based system to offer continuity in the communication of a given mobile node with a remote proxy \cite{GhiniJSS}. Through this multihoming solution the mobile node is enabled to automatically switch from one network to another (e.g., from UMTS to WiFi) without interrupting the end-to-end communication with the correspondent node, and transparently to the human user. This is possible using a proxy which hides the handover from a network to another (and hence, also the change of IP address for the mobile node). The use of a proxy placed in the Internet solves several problems concerned with the identification of a mobile node trying to get access to Internet services in a smart shire. In fact, a message originated at a mobile node can arrive in the Internet following several paths and possibly with different IP addresses.  This can be a problem, since in the Internet the IP address usually identifies the sender of a message \cite{Ferretti2016390}. To cope with this issue, the proxy can act on behalf of the user by relaying requests to Internet services or correspondent nodes. This way, at the destination, the sender of the message appears to be the proxy itself.
This solution has the advantage of avoiding the need for modifications at the Internet services or correspondent nodes. This is in accordance to the need for cheap solutions to be employed in smart shires.

We performed a simulation assessment where we compare three multihoming approaches, i.e.~ABPS \cite{GhiniJSS}, MIPv6 \cite{mipv6} and LISP \cite{lisp}. Results show the effectiveness of our proposal.

The remainder of this paper is organized as follows. Section \ref{sec:s_shires} describes the main aspects concerned with smart shires.
Section \ref{sec:archi} discusses on the principal wireless communication solutions and protocols to be used to enable the development of viable and smart communication services in smart shires. Section \ref{sec:abps} presents ABPS, our proposal for the support of multihoming. Section \ref{sec:eval} shows a simulative evaluation we performed. Finally, Section \ref{sec:conc} provides some concluding remarks.

\section{Smart Shires}\label{sec:s_shires}

A smart shire can be through as a middleware of services, deployed at different levels, involving different technologies and different parties, together with the applications that can be deployed on top of this platform. The technologies and software involved in the development of a smart shire should be based on the social goals that should be reached. Applications are related to the deployment of improved Internet access, access to information, proximity-based applications (e.g., proximity-based social networking, advertisements for by-passers, local exchange of information), traffic control, security and public safety support, smart metering, smart agriculture and animal farming, services to municipalities such as smart traffic management systems, data collection through environmental sensors for monitoring resources and facilities, smart eHealth, security and emergencies. In order to being ready for critical events, such as natural disasters (e.g., earthquakes, floods, fires) sensor networks can be proficiently deployed in the territories (e.g., riverbanks, woods), to be used used in smart services. Other applications might refer to the optimization of the supply chain processes in the rural territory, in conjunction with urban regions in the smart territory \cite{smartshires}.

The middleware must be able to collect and disseminate contents among services, public and private organizations, users, sensors. Smart shires should make use of crowd-sensed and crowd-sourced data to generate information to be used within services. We can assume to have cheap sensors, forming a cloud of things, deployed all over the countrysides. Thus, the use of open data is a promising option to promote data sharing. Then, services using these data must guarantee scalability. As a consequence, computing systems such as cloud or fog architectures must enter into the picture.

Finally, communication infrastructures include networks such as 5G and broadband, WiFi, WiMax networks, but also ``infrastructure-less'' networking approaches (e.g., MANETs, VANETs, Internet of Things) and sensor networks.

\section{Ad-hoc, Mesh Networks and Multihoming}\label{sec:archi}

The idea of identifying novel paradigms for an optimal organization of devices in a smart shire might appear rather ambitious and complicated. However, most of the technologies needed to achieve these target paradigms are already available; they just need to integrate with each other. These technologies include opportunistic, self-organizing networks, multihoming techniques, Device-to-Device (D2D), wireless ad-hoc and mesh networks. All these solutions should optimize the use of all the wireless communication technologies available in the territory, such as cellular, WiFi, WiMax networks.

\begin{figure}[t]
\centering
 \includegraphics[width=\linewidth]{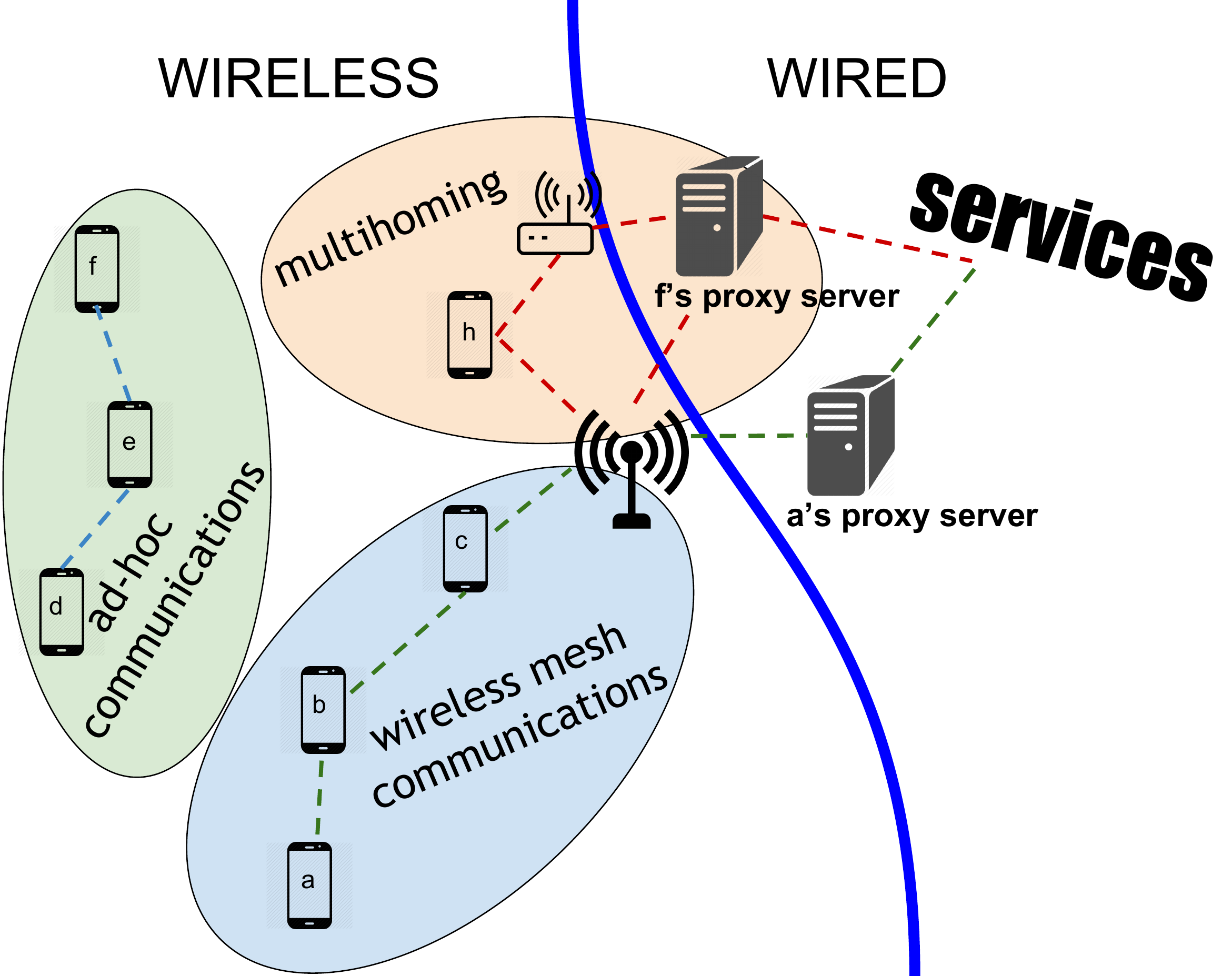}
\caption{Multipoint and multihoming communications.}
\label{fig:mhop}
\end{figure}

Figure \ref{fig:mhop} summarizes the different types of communications that may arise in the considered smart shire scenarios. In the figure, nodes $d, e$ and $f$ perform an ad-hoc communication, without the intervention of a network infrastructure. Node $a$ is able to connect to Internet services via a wireless mesh network, that exploits a multihop communication to let messages reaching a network infrastructure \cite{Alotaibi2012940}. Finally, $h$ connects to Internet services thanks to the use of multihoming, that allows utilizing its multiple network interface cards concurrently, in a seamless way \cite{Ferretti2016390}.

Sensors are relatively cheap in terms of costs. Thus, their deployment in a countryside is feasible. Problems may arise to interconnect these sensors to form a sensor network and to make them communicate with the intelligent service placed in the Internet. This requires the use of smart services employing D2D, wireless ad-hoc, mesh networks and multihoming techniques.

Ad-hoc communications are regarded as promising technologies to provide low-power, high-data rate and low-latency services between end-users in the future wireless networks \cite{Yilmaz}. Depending on the used network technologies, we might refer to D2D cellular networks, which is a prominent technology for 5G networks \cite{Asadi}, rather than other forms of mobile mesh or ad-hoc networks. D2D is defined as direct communication between two mobile users using a cellular interface, without traversing the Base Station or core network. WiFi technologies enable the creation of Mobile Ad-hoc NETworks (MANETs), or Vehicular Ad-hoc NETworks (VANETs) when the nodes are machines, when mobile nodes are interconnected in an ad-hoc fashion, or so called wireless mesh networks when the network includes access points that allows routing messages to the Internet, and vice versa \cite{Alotaibi2012940}. Ad-hoc communications might serve to run (at the higher levels) proximity based applications, exploit multihop facilities to discover other nodes/services and/or access points able to route messages to the Internet.

An important feature to be provided is the ability to allow a mobile node, having multiple wireless network interfaces, to change network points of attachment (handover) without disrupting existing connections \cite{Ferretti2016390}. 

\subsection{The Roles of Devices} 

The main objective to pursue is a novel distributed architecture that enables the dissemination of open data, where all available devices are dynamically and adaptively configured depending on: i) the devices themselves, ii) the environment in which they are deployed, and iii) the other devices (their characteristics) and the interactions they have \cite{gda-emergency-11}. The architecture must provide configuration protocols to automatically organize all devices, identify algorithms and mechanisms for the simultaneous and adaptive use of different communication networks in an opportunistic fashion \cite{Sambo2014}. Moreover, it is necessary to develop protocols for the organization and interaction between devices. In other terms, mechanisms for the efficient resource and data sharing are needed. 

In general, we foresee that sensors will be available to build a cloud of things, but even devices of mobile users may decide to participate to the services, by enabling message relay functionalities and/or offering its sensors to crowdsensing services. Thus, devices can be configured depending on the running applications, devices' energy and computation constraints, and on the users willingness to participate to the smart shire distributed systems. From a communication standpoint, we can thus identify different communication scenarios for a given (mobile) node, equipped with multiple network interface cards:
\begin{itemize}
 \item A single network interface card is available; in this case, if a traditional network infrastructure is present, the node might prefer connecting to that base station / access point directly. Otherwise, the node might connect to a wireless mesh network or an ad-hoc network, if present.
 \item Multiple network interface cards are available (commonly, a cellular and a WiFi network interface card); in this case, some optimized configuration protocol should be executed to understand what is the best choice, based on the user preferences and devices available in the proximity of the considered node. An option might be that of exploiting one network to connect to the Internet and the other one to perform mesh / ad-hoc communications with mobile nodes in the same geographical area.  
\end{itemize}

It is clear that several possible situations may occur and there is a strong need for smart strategies to manage all possible networking opportunities. 
These strategies provide the substrate to build smart communication services at the higher levels.

\subsection{The Need for Proxies}
In general, a proxy is a software entity in charge of routing data/information towards a given host, typically in compliance with an application level protocol. The proxy represents the software entity which is in charge of managing the mobility of a user. It might accomplish several tasks, such as node localization, user identification, packet relaying, etc. This allows the mobility management service to be deployed with no impact on the network infrastructure.

The idea is that in smart shire scenario, a mobile node can traverse the wireless network using multihoming (i.e., multiple networks in a seamless way) and/or ad-hoc communications in a mesh network. Thus, a message originated at a mobile node can arrive in the Internet following several paths and possibly with different IP addresses. This is a problem, since in the Internet the network (IP) address usually identifies the sender of a message (and this is in contrast with the scenario we are discussing) \cite{Ferretti2016390}. A smart proxy can identify the originator of a message, regardless on the path the message followed to reach it. Then, the proxy can act on behalf of the user by relaying requests to Internet services or correspondent nodes. This way, at the destination, it appear as if the sender of the message is the proxy itself. This solution has the advantage of avoiding the need for modifications at the Internet services or correspondent nodes. Moreover, the use of a proxy typically overcomes the presence of firewalls and NAT systems \cite{Ferretti2016390}.

To factually implement such functionalities, this proxy (hereinafter referred as ``proxy server'') can be coupled with another software component, running inside the mobile node (hereinafter referred as ``proxy client'') \cite{Ferretti2016390}. The idea of exploiting such a pair of software components allows client applications to be unaware of the fact that at the lower levels, the mobile node experiences network reconfigurations and that exchanged messages may travel through multihop and multipath wireless communications. The mobile node only needs to be configured to use the proxy client running on the local host.

It is important to mention that since each mobile node requires a correspondent proxy server, for the sake of scalability proxy servers can be placed in cloud or (in the opposite case) fog computing infrastructures. Only mobile devices (e.g., users' phones) need a specific instance of a proxy server to support their mobility. Conversely, other static sensors do not have such a requirement; they require a preliminary setup phase, without dynamic and frequent reconfigurations.

\subsection{Wireless Ad-Hoc Communications: Viable Data Dissemination in Smart Shires}

We already mentioned that ad-hoc communications represent an important tool to distribute and deliver messages in smart shires. In fact, 
in proximity based applications devices being close to each other can activate direct links and bypass the base station or access point by either using cellular D2D communications \cite{Asadi}, ad-hoc Wi-Fi based solutions such as MANETs \cite{Boukerche20113032} or even VANETs \cite{Palazzi:2010}. An example of these kinds of interaction is reported in Figure \ref{fig:mhop}, where nodes $d, e, f$ interact in an ad-hoc communication; if $d$ and $f$ are not able to communicate directly, since their transmission range is limited, then $e$ might relay messages coming from $d$ towards $f$ and vice versa.

\subsubsection{Reaching an Area of Interest or a Specific Destination}
Depending on the application scenario a message might have to be sent from a device towards a certain physical area of the territory (e.g., the device is looking for a sensor to understand weather conditions) or towards a specific destination, or the message ``is looking for an access point'', trying to reach an Internet host or service.
In this case, a MANET-like routing protocol can be employed. A plethora of works is available in the literature, which are discussed in several survey papers, e.g., \cite{Alotaibi2012940,Boukerche20113032}. 

\subsubsection{Broadcast in a given Area of Interest}
In other cases, applications might require to broadcast a message; this is the case for alert messages, critical situations, or more simply general information or advertisements. To this aim, at the lower, datalink level, an efficient broadcast scheme might be employed that spread messages across devices, trying to avoid transmissions' collisions \cite{Hong2014}. This provides a useful scheme to be employed at higher levels and disseminate requests and contents. The broadcast might have a sort of Time-To-Live counter that limits the number of retransmissions, or some information related to the geographical area of interest acting as a boundary, so that when a message leaves its area of interest, it is not retransmitted anymore.

A plethora of approaches is available in literature \cite{Palazzi:2010}. The rationale behind these schemes is to regulate the nodes wireless transmissions so as to avoid congestions, redundant multihop transmissions and to augment reliability and efficiency. Typically, a backoff mechanism is used that reduces the frequency of message retransmissions when congestion is causing collisions. In fact, if no intelligence is applied to the multihop broadcasting scheme, any node in the network would simply relay every received message. The consequent explosion of transmitted messages would lead to high congestion, collisions and delays, a phenomenon known as the ``broadcast storm problem'' \cite{bani}.
Thus, as soon as a node receives a message that needs to be broadcast in a certain area, it relays the message only when no other ``more suitable'' nodes are available to relay that message. 
Nodes thus utilize an efficient priority scheme to choose the next-hop forwarder based on the distance from the previous sender and on the expected transmission range. Farther nodes have higher priority than others when forwarding received messages. Nodes’ priorities to forward a message are determined by assigning different waiting times from the reception of the message to the time at which they will try to forward it, similarly to classical backoff mechanisms in IEEE 802.11 medium-access control (MAC) protocols \cite{Palazzi:2010}.
This way, redundant transmissions (and message propagation delays) are reduced. In \cite{smartshires}, we show that such a kind of ``priority-based broadcast'' can effectively spread contents in ad-hoc networks employed in smart shire scenarios.

\section{Access to the Internet: Multihoming Through Always Best Packet Switching}\label{sec:abps}

This section deals with issues concerned with multihoming. 
Figure \ref{fig:mhop} shows an example where node $h$ has two different infrastructured networks available; thus, it can exploit a multipath communication, by selecting the most suitable access network dynamically (multihoming). Changes of networks must be accomplished in a fast and seamless way, in order to preserve a good Quality of Service perceived at the application level.

A plethora of works has been proposed in the literature about multihoming mobility architectures; the reader can refer to \cite{Ferretti2016390} for a wide survey on this topic. Our solution, termed  Always Best Packet Switching (ABPS), is a cross-layer technique that aims at limiting the need for modifying the network infrastructure. This is accomplished by resorting to an instance of the previously mentioned proxy server \cite{FerrettiG09,GhiniJSS}. This solution splits the communications in two consecutive paths (see Figure \ref{fig:mhop}): 
\begin{enumerate}
 \item a (multipath) communication from the mobile node to the proxy server;
 \item the Internet-based traditional communication between the proxy server and the correspondent node/service.
\end{enumerate}

The current implementation of ABPS supports SIP/RTP applications, such as VoIP and Video on Demand (VoD) \cite{GhiniJSS}, and HTTP-based Web 2.0 services \cite{FerrettiG09}. This approach enables the transmission of each datagram through the most appropriate network interface card among those which are active at the mobile node.

ABPS takes into consideration QoS metrics to identify the best network interface to use at any given moment. A cross-layer technique is employed to monitor all the concurrent network interfaces which are available, their performances and those that become active (or inactive).
A software tool is executed at the data-link layer that provides the mobile node with information about successful (unsuccessful) datagram reception at the access point \cite{GhiniJSS}. This information can be delivered to software modules working at higher (i.e.~application) layers of the protocol stack. This way, it is possible to dynamically managing network interfaces and performing vertical handovers.

ABPS uses a specialized version of the pair of proxies mentioned in the previous section. In fact, the implemented proxies are session/application-compliant (e.g., SIP) proxy servers. They are responsible for managing the application-layer data flows, enabling their transmission through different networks, on a per-packet basis. Moreover, each proxy adds a digital signature in the packet so that the proxy server may identify the sender, in spite of its possible different IP addresses due to the multipath communications (e.g., see $h$ node in Figure \ref{fig:mhop}) and of regardless of the fact that a message has traveled across multihop communications (e.g., see $a$ node in Figure \ref{fig:mhop}). A detailed description of these software components can be found in \cite{GhiniJSS}.

\subsection{Advantages of ABPS}\label{sec:advantages}
The use of ABPS as a multihoming strategy in smart shires has three main advantages.

First, ABPS does not require any modification to the network infrastructure. It has been designed to work in whatever IP-based network. Conversely, other main proposals for multihoming, e.g., Mobile IP version 6 (MIPv6) \cite{mipv6}, Location/ID Separation Protocol (LISP) \cite{lisp} need some specific software component running in the access network that the mobile node is connecting to, e.g., a home agent for MIPv6, ingress and egress tunnel routers for LISP. Not only, these solutions require some additional software components also at the correspondent node side. For example, in MIPv6 the correspondent node should be able to perform a binding update when the mobile node has a novel care of address; in LISP, the correspondent node needs to be in a network with a LISP-enabled router. It is worth noting that in a countryside it would be difficult to deploy an evolved network infrastructure with additional software components. Rather, a simpler deployment approach, such as that of ABPS, should be preferred.

Second, ABPS reduces the downtimes, i.e, periods when a system is unavailable, thus fostering its use to support highly interactive applications \cite{GhiniJSS}. This is obtained through the concurrent management and configuration of multiple network interfaces, that can be used at any time. Thus, as soon as a network becomes unreachable, ABPS is able to switch to the another one, if a connection is available, without the need to configure the connection at that time (since it is already configured).

Third, being based on the existing network infrastructure, ABPS does not need the (costly) dynamic configuration of communication paths between software entities; only the proxies have to be configured during the system initialization. Conversely, MIPv6 based protocols require a set of control operations and paths setup, e.g., during the return routability procedure performed when a mobile node changes its access network, that take some time. In LISP, a configuration protocol is needed when the correspondent node is outside a LISP-enabled sub-network.

\section{Performance Evaluation}\label{sec:eval}

\begin{figure*}[t]
\centering
 \includegraphics[width=\linewidth]{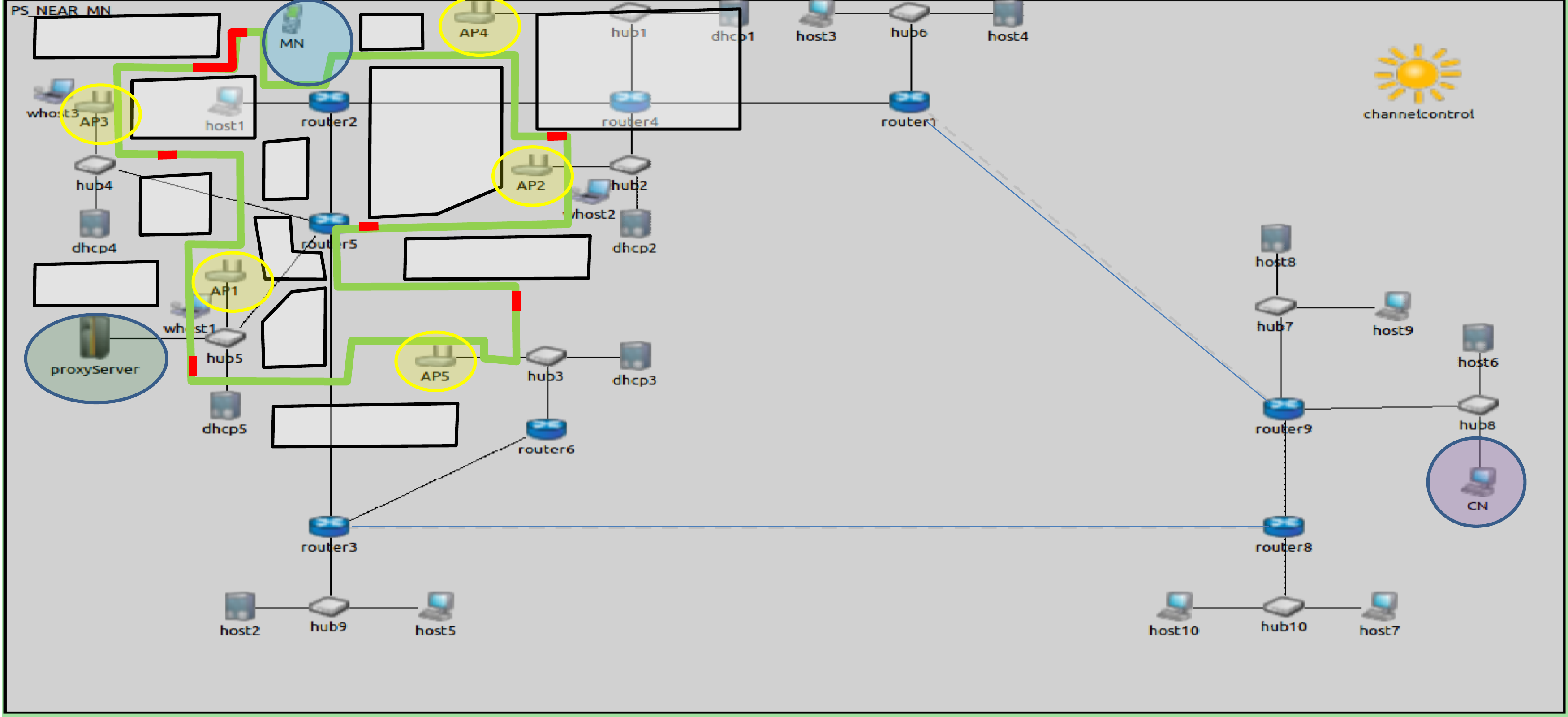}
\caption{Simulation scenario, OMNet++ simulation model: green/red segments represent the user path, the access points are highlighted with yellow circles and the ABPS proxy server is highlighted with a blue ellipse. Black polygons, in the simulated area, represent the obstacles.}
\label{fig:scenario}
\vspace{-0.4cm}
\end{figure*}

In order to assess the viability of a multihoming architecture in a smart shire, we created a simulative scenario using the OMNet++ simulator (version 4.3.1) and the INET framework. A mobile user was equipped with a mobile device able to exploit two network interface cards concurrently (for the sake of simplicity and the need to easily simulate network disconnections, these were configured as two WiFi cards). The user moves in a path, as depicted in Figure \ref{fig:scenario}, with several access points of different networks, i.e.~each access point belongs to a different WLAN. Obstacles obstruct wireless communications; thus, while moving the user might have multiple available networks, a single network, or even no network connectivity. In the figure, access points are emphasized with yellow circles. The user itinerary is the depicted green/red path, where green means that a wireless connectivity is available for the user, while red means that an handover occurs. During the path, there is a situation when the user has no connectivity at all while moving (the longest red path in the figure).

In Figure \ref{fig:scenario}, the blue circled entity represents the ABPS proxy server that interacts with the mobile node. The choice of placing the ABPS proxy server geographically near the mobile node is in accordance with the mentioned fog computing system architecture, where computing resources for the support of smart services are placed at the edge of the network.

We tested three different approaches for mobility management: i.e., our mentioned ABPS \cite{GhiniJSS}, MIPv6 \cite{mipv6} and LISP \cite{lisp}. Based on the issues mentioned in Section \ref{sec:abps}, we focused on the downtime (i.e., time of service unavailability)
occurring during the handovers. In this case, the downtime is measured as the time elapsed between the first failed transmission of a sent packet, due to a wireless network disconnection and the time when that retransmitted packet (or one of its successors) is finally delivered to the destination (since a wireless network connection becomes available). All the results reported in this section are averages of multiple independent runs. In all cases, the confidence intervals have been calculated but not reported for readability.

\begin{figure}[t]
\centering
 \includegraphics[width=\linewidth]{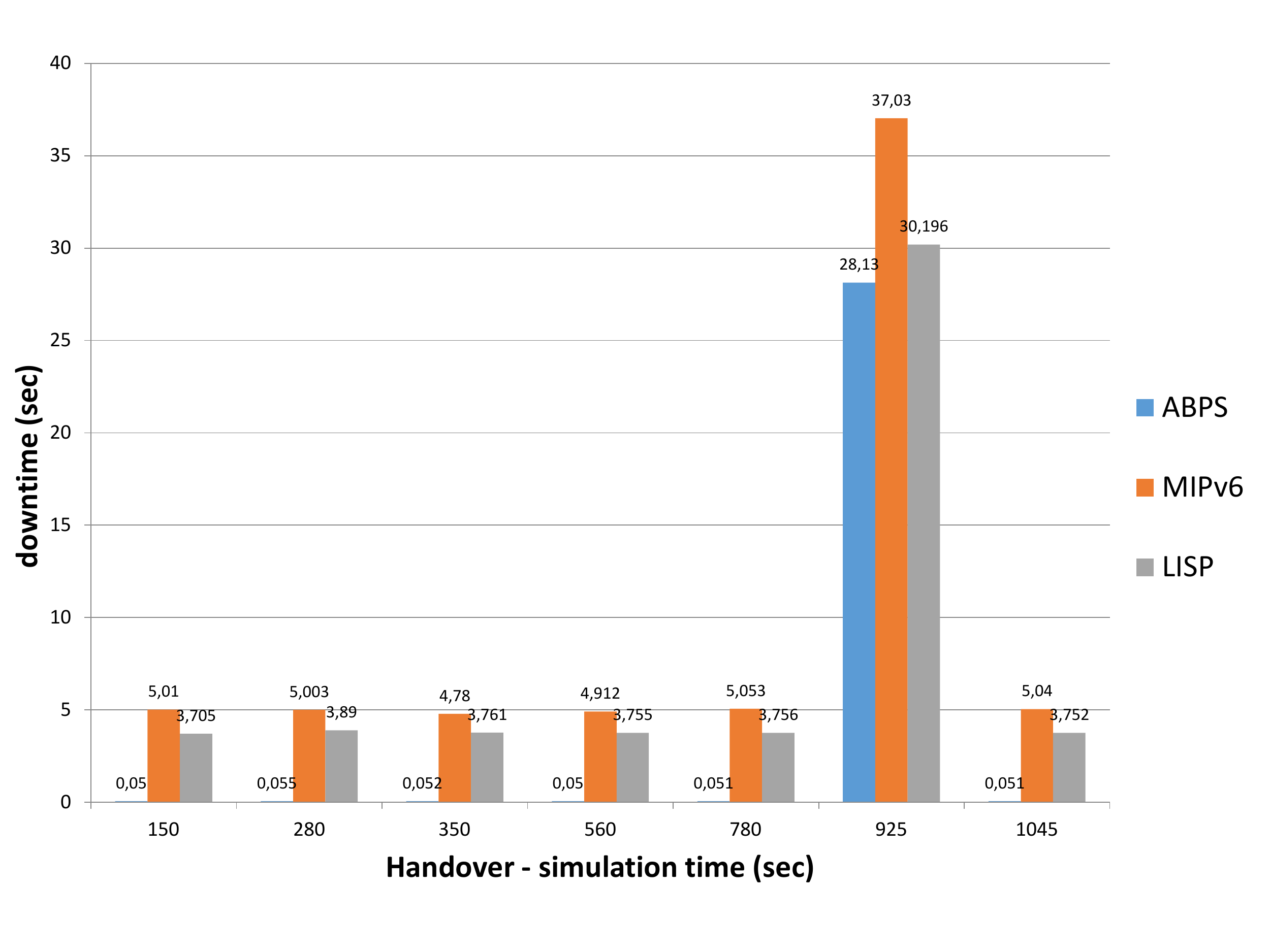}
\vspace{-0.5cm}
\caption{Measured handover downtime in the simulated scenario.}
\label{fig:res}
\vspace{-0.4cm}
\end{figure}

Figure \ref{fig:res} shows the downtime measured using the three mentioned approaches. It is possible to notice that in most cases, with ABPS the handover requires $\sim0.05$sec to be accomplished, since no reconfigurations are needed. Conversely, reconfiguration times are noticeable with MIPv6 ($>5$sec) and LISP ($>3$sec). As concerns the situation of network unavailability (i.e., simulation time $925$sec), ABPS shows a lower downtime w.r.t.~MIPv6 and LISP. 

These results demonstrate the effectiveness of the proposed approach. This means that, multihoming is a principal building block for the definition of a smart shire architecture. In other words, we envision a system in which the different user devices can be organized for a smart use of their network interfaces, in order to maximize the connectivity in a (typically) limited and intermittent connectivity environment.

\section{Conclusion}\label{sec:conc}

In this paper, we described network issues and practical solutions to deploy smart services over countrysides, in order to build effective smart shires. Multihop and multipath approaches are to be used in order to guarantee network connectivity in very diverse conditions, where a network infrastructure can be present only in certain portions of the considered area.

We put the focus on ABPS, a multihoming approach for the support of node mobility. A plus of this approach is that it does not require any modification to the current structure of the Internet; this is in accordance to the need of cheap solutions for the support of smart shires. We described its main characteristics and showed that it can outperform other typical approaches. This demonstrates that smart communication solutions can be devised and effectively deployed in countrysides in order to build smart shire services.

\small{
\bibliographystyle{abbrv}
\bibliography{paper}  
}

\end{document}